\documentclass[aps,twocolumn,superscriptaddress,showpacs,nofootinbib]{revtex4-2}

\usepackage[dvips]{graphicx}
\usepackage{amsmath,amssymb,amsthm,mathrsfs,amsfonts,dsfont}
\usepackage{siunitx}
\usepackage{subfigure, epsfig}
\usepackage{physics}
\usepackage{bm}
\usepackage{enumerate}
\usepackage{color}
\usepackage{qcircuit}
\usepackage[normalem]{ulem} 
\usepackage{epstopdf}
\usepackage{comment}
\usepackage{diagbox}
\usepackage{multirow}    
\usepackage{array}     
\usepackage{enumitem} 
\usepackage{booktabs,colortbl} 
\usepackage{rotating}
\usepackage{ulem}
\usepackage{longtable}
\usepackage{setspace}
\usepackage{upgreek}
\usepackage[ruled,vlined]{algorithm2e}
\usepackage[bookmarks=false,colorlinks,citecolor=blue,linkcolor=blue,anchorcolor=blue,urlcolor=blue]{hyperref}

\renewcommand{\i}{\mathrm{i}}

\begin{document}
\preprint{APS/123-QED}

\title{Experimental Phase-Matching Quantum Cryptographic Conferencing in Symmetric and Asymmetric Fiber Channels}

\author{Mi Zou}
\affiliation{Hefei National Laboratory, University of Science and Technology of China, Hefei 230088, China}

\author{Bin-Chen Li}
\affiliation{Hefei National Laboratory, University of Science and Technology of China, Hefei 230088, China}
\affiliation{Hefei National Research Center for Physical Sciences at the Microscale and School of Physical Sciences, University of Science and Technology of China, Hefei 230026, China}
\affiliation{CAS Center for Excellence in Quantum Information and Quantum Physics, University of Science and Technology of China, Hefei 230026, China}

\author{Shuai Zhao}
\affiliation{School of Cyberspace, Hangzhou Dianzi University, Hangzhou 310018, China}
\affiliation{Pinghu Digital Technology Innovation Research Institute, Hangzhou Dianzi University, Pinghu,314200,China}

\author{Yingqiu Mao}
\affiliation{Hefei National Research Center for Physical Sciences at the Microscale and School of Physical Sciences, University of Science and Technology of China, Hefei 230026, China}
\affiliation{Shanghai Research Center for Quantum Science and CAS Center for Excellence in Quantum Information and Quantum Physics, University of Science and Technology of China, Shanghai 201315, China}

\author{Dandan Qin}
\affiliation{Hefei National Laboratory, University of Science and Technology of China, Hefei 230088, China}
\affiliation{Hefei National Research Center for Physical Sciences at the Microscale and School of Physical Sciences, University of Science and Technology of China, Hefei 230026, China}
\affiliation{CAS Center for Excellence in Quantum Information and Quantum Physics, University of Science and Technology of China, Hefei 230026, China}

\author{Xiao Jiang}
\affiliation{Hefei National Laboratory, University of Science and Technology of China, Hefei 230088, China}
\affiliation{Hefei National Research Center for Physical Sciences at the Microscale and School of Physical Sciences, University of Science and Technology of China, Hefei 230026, China}
\affiliation{CAS Center for Excellence in Quantum Information and Quantum Physics, University of Science and Technology of China, Hefei 230026, China}

\author{Teng-Yun Chen}
\email{tychen@ustc.edu.cn}
\affiliation{Hefei National Laboratory, University of Science and Technology of China, Hefei 230088, China}
\affiliation{Hefei National Research Center for Physical Sciences at the Microscale and School of Physical Sciences, University of Science and Technology of China, Hefei 230026, China}
\affiliation{CAS Center for Excellence in Quantum Information and Quantum Physics, University of Science and Technology of China, Hefei 230026, China}

\author{Jian-Wei Pan}
\email{pan@ustc.edu.cn}
\affiliation{Hefei National Laboratory, University of Science and Technology of China, Hefei 230088, China}
\affiliation{Hefei National Research Center for Physical Sciences at the Microscale and School of Physical Sciences, University of Science and Technology of China, Hefei 230026, China}
\affiliation{CAS Center for Excellence in Quantum Information and Quantum Physics, University of Science and Technology of China, Hefei 230026, China}

\begin{abstract}
Quantum cryptographic conferencing (QCC) allows multiple parties to establish common secure keys in quantum networks with information-theoretic security. However, the secure transmission distances of current QCC implementations are still limited to the metropolitan areas. Here, we experimentally demonstrate the three-intensity phase-matching (PM) QCC protocol considering finite-size effects by employing frequency-locking and phase-tracking techniques for three parties. The key distribution capability of the PM QCC protocol is demonstrated in the symmetric fiber channels with the distance from each party to the measurement site up to $\SI{100}{\km}$. The network adaptability of the PM QCC protocol is demonstrated in asymmetric fiber channels used to simulate fiber channel configurations in real networks. Thus, the feasibility of applying the PM QCC protocol to practical intercity quantum networks with both symmetric and asymmetric channels is verified. 

\end{abstract}

\maketitle

\section{Introduction}

Quantum networks aim to achieve information processing tasks beyond the capabilities of classical networks \cite{NKimble2008,SWehner2018, JoPAMaTNokkala2024}, with quantum key distribution (QKD) being a key application that enables secure key generation between remote parties \cite{RMPScarani2009,RMPXu2020}. Various QKD networks, featuring different topologies and scales, have been developed to enhance the practical use of quantum information \cite{NJoPPeev2009, OEChen2010, OESasaki2011, PRXTang2016, SAJoshi2020, NChen2021, PRAZhong2022}. When multiple parties in a network intend to generate a common secure key, the approach using bipartite QKD protocols requires either one party to generate independent identical secure key pairs with each of the remaining parties individually or adjacent parties to generate their own independent identical secure key pairs between themselves. These pairwise keys are then converted into a multi-party common key via classical XOR operations. An alternative, more concise and efficient approach is to directly generate a multi-party common secure key by leveraging multipartite entanglement \cite{QICChen2007, PRLFu2015, NJoPEpping2017, NJoPGrasselli2018, NJoPGrasselli2019, AQTMurta2020, PRAZhao2020, IACao2021, NJoPCao2021, EHua2022, PRACarrara2023,CPXie2024, RoPiPLu2025}.

\begin{figure}[b]
	\includegraphics[width=0.9\linewidth]{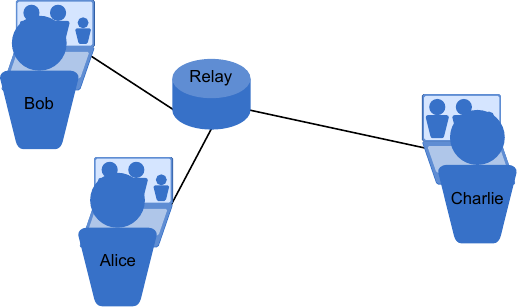}
	\caption{Three parties in a star-type quantum network are conducting an encrypted conference using the common key generated by the QCC protocol. The relay can be an untrusted entanglement source or measurement site, and the distance from each party to the relay may be different.}
\label{fig:setting} 
\end{figure}

Figure~\ref{fig:setting} depicts a star-type quantum network application facilitating an encrypted conference among three parties using the quantum cryptographic conferencing (QCC) protocol, which generates secure common keys. This protocol can utilize either an entanglement-based approach that distributes the genuine multi-party entangled state \cite{QICChen2007, NJoPEpping2017, NJoPGrasselli2018} or time-reversal methods that postselect the Greenberger-Horne-Zeilinger (GHZ)\cite{PRLFu2015, PRAZhao2020, CPXie2024} or W states \cite{NJoPGrasselli2019, PRACarrara2023}. While the entanglement-based QCC is limited by the fragility of multi-party entangled states \cite{SAProietti2021, nQIPickston2023}, the time-reversal QCC, employing weak coherent states, offers greater practicality and is resistant to detector side-channel attacks, making it measurement-device-independent (MDI). Recent implementations of the MDI QCC protocol, based on polarization encoding, have achieved secure key rates constrained by inherent transmittance limitations \cite{PRLYang2024,PRLDu2025}. The proposed phase-matching QCC (PM QCC) protocol \cite{PRAZhao2020} enhances the key rate from $O(\eta^M)$ to $O(\eta^{M-1})$ with respect to the transmittance $\eta$ from each party to the measurement site and the number of parties $M$, allowing for longer secure transmission distances. However, the original PM QCC assumes symmetric channels, while real networks often exhibit asymmetric channels, necessitating loss compensation methods to achieve symmetry.

In this work, we extend the security proof of the original PM QCC protocol to asymmetric channels, an issue of widespread concern in QKD protocols \cite{PRLLiu2019, PRXWang2019, PRAZhou2019, PRAHu2019, NJoPGrasselli2019a, NJoPWang2020, nQIZhong2021,OZhu2024}, and refine the protocol into a three-intensity scheme, which simplifies intensity modulation. In this way, the PM QCC protocol can be implemented in asymmetric channels by optimizing the transmitted light intensity of each party instead of compensating for the loss. We experimentally demonstrate the three-intensity PM QCC protocol considering finite-size effects \cite{PRAZhang2017} by employing frequency-locking and phase-tracking techniques for three parties. The refined PM QCC protocol is implemented in symmetric channels with the distance from each party to the measurement site increasing from $\SI{25}{\km}$ to $\SI{100}{\km}$ in intervals of $\SI{25}{\km}$ and in five asymmetric fiber channel configurations chosen from an arrangement of $\SI{25}{\km}$, $\SI{50}{\km}$ and $\SI{75}{\km}$. These results promote the application of QCC protocols in real quantum networks.

\section{Protocol}
The three-intensity PM QCC protocol begins by Alice preparing the coherent state $\ket{e^{\i\varphi_A}\sqrt{\mu_A}}$ for each round. The modulated intensity $\mu_A$ is randomly chosen from one signal-state intensity and two decoy-state intensities denoted as $\{ \mu_a, \nu_a, \omega_a \}$ with probabilities $\{ p_\mu, p_\nu, p_\omega\}$, where $\omega_a<\nu_a<\mu_a$, $p_\mu+ p_\nu+p_\omega=1$. The modulated phase $\varphi_A=\phi_a+\pi k_a$, where $k_a \in\{0, 1\}$ is one random bit, and $\phi_a$ is randomly chosen from $ \{ 0,\frac{2\pi}{D},\frac{4\pi}{D},...,\frac{2\pi(D-1)}{D} \} $. Here, $D$ is the number of phase slices. Similarly, Bob (Charlie) prepares the coherent state $\ket{e^{\i\varphi_B}\sqrt{\mu_B}}$ ($\ket{e^{\i\varphi_C}\sqrt{\mu_C}}$) with the intensity $\mu_B \in \{\mu_b,\nu_b,\omega_b\}$ ($\mu_C\in \{\mu_c,\nu_c,\omega_c\}$) and the phase $\varphi_B=\phi_b+\pi k_b$ ($\varphi_C=\phi_c+\pi k_c$). Here, $k_b$ ($k_c$) is a random bit similar to $k_a$, and $\phi_b$ ($\phi_c$) is a random phase similar to  $\phi_a$. They then send their coherent states to the untrusted measurement site, Eve. An honest Eve splits Alice's coherent state into two parts to interfere with Bob and Charlie's coherent states, respectively. Thus, two measurement branches are created, followed by two single photon detectors in each branch. Then Eve records the successful detection event, i.e., only one detector click for each measurement branch in that round. 

After repeating the above procedure for many rounds, Eve announces all the successful detection events, Alice, Bob, and Charlie announce their random phases and intensities and perform the sifting step to obtain raw keys. They retain or flip their random bits only when the following phase-matching conditions are satisfied: $|\phi_a-\phi_b|=0~\text{or}~\pi$, $|\phi_a-\phi_c|=0~\text{or}~\pi$; and the intensities of coherent states they prepared are $\{ \mu_a,\mu_b,\mu_c \}$, $ \{ \nu_a,\nu_b,\nu_c \}$, or $\{ \omega_a,\omega_b,\omega_c \}$, respectively. The secure keys are obtained by performing parameter estimation and key distillation. The security proof of the protocol is realized via entanglement-based method \cite{PRAZhao2020} and source replacement method \cite{AQTHuang2024} (see Appendix \ref{sec:sa} for more details). The key rate of the three-intensity PM QCC protocol for three parties can be expressed as 
\begin{equation} 
R = \left( \frac{2}{D} \right)^2 Q_\mu \left[1-fh(E^\text{max}_Z)-h(E^U_X)\right],
\end{equation}
where $(2/D)^2$ is the prefactor induced by phase post-selection, $Q_{\mu}$ is the gain of signal state, $f$ is the error correction efficiency, $h(x)=-x \log_2(x)-(1-x)\log_2(1-x)$ is the binary entropy function.  $E^\text{max}_Z$ is the maximum value between $E^{AB}_Z$ and $E^{AC}_Z$, where $E^{AB}_Z$  ($E^{AC}_Z$) is the quantum bit error rate between Alice and Bob (Charlie), which can be directly obtained from the experimental results. $E^U_X$ is the upper bound of the phase error rate, which can be estimated by decoy-state method given in Appendix \ref{sec:sa}.

\section{Experiment}
\begin{figure*}[t]
	\centering 
	\includegraphics[width=\linewidth]{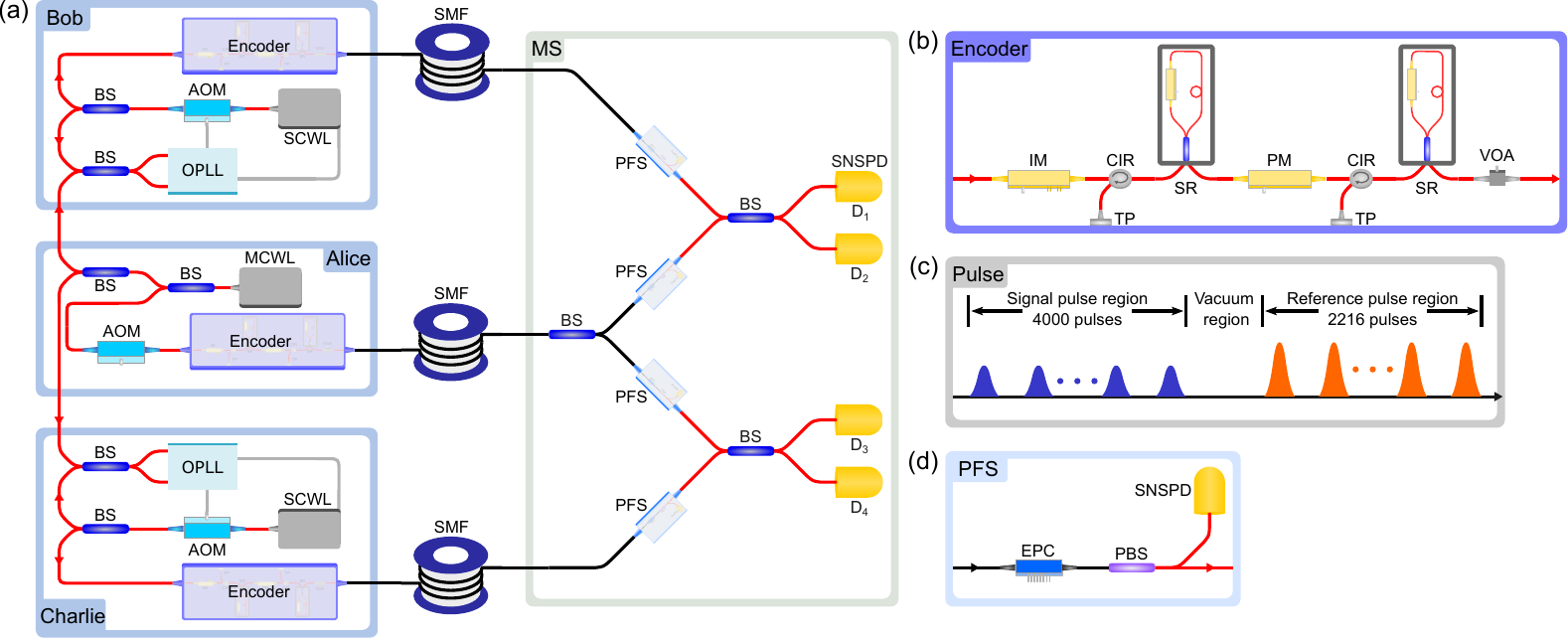}
	\caption{Experimental setup of PM QCC protocol for three parties. (a) Layout: Three parties Alice, Bob, and Charlie are connected to the measurement site (MS) through single-mode fiber (SMF). Bob and Charlie's continuous-wave lasers (CWLs) act as slave CWLs (SCWLs), frequency locked to Alice's master CWL (MCWL) using heterodyne optical phase-locked loop (OPLL) technology with acousto-optic modulator (AOM) enhancing feedback bandwidth. Alice then uses an AOM to align the frequency of remaining MCWL beam with those of Bob and Charlie. The three parties have the same encoder for intensity and phase modulation of coherent light pulses. At the MS, Alice's coherent state is split by beam splitter (BS) into two parts, which interfere with the coherent states of Bob and Charlie in the other two BS after passing through the polarisation feedback system (PFS). The interference results are then detected by four channels of superconducting nanowire single-photon detector (SNSPD). (b) Encoder: The encoder consists of an intensity modulator(IM), two Sagnac rings (SRs), two circulators (CIRs) in front of two SRs, a phase modulator (PM) between the two SRs, and a variable optical attenuator (VOA). A test port (TP) is reserved for each circulator. (c) Pulse: The encoder generates 62500 pulses per $\SI{100}{\us}$ with 10 repeating regions excluding the pulses used for time calibration. Each repeating region consists of a reference pulse region, a vacuum region, and a signal pulse region. There are 2216 pulses in each reference pulse region and 4000 pulses in each signal pulse region. (d) PFS: The PFS is used to align the polarization of coherent light pulses. Each PFS is composed of an electric polarization controller (EPC), a polarization beamsplitter (PBS), and one channel of the detector.}
\label{fig:setup}
\end{figure*}
The three-intensity PM QCC protocol is implemented with the setup shown in Fig.~\ref{fig:setup}.  All three parties have coherent light sources (CLSs) and encoders for coherent state preparation. To establish a stable phase reference between three parties, they first need to ensure that their CLSs have the same frequency. To this end, the continuous-wave laser (CWL) of Alice is used as the master CWL (MCWL) to send out two beams of frequency reference light. Bob and Charlie receive two frequency reference beams from Alice and lock their slave CWLs to them. They use heterodyne optical phase-locked loop technology \cite{NPMinder2019} to make a stable  $\SI{50}{\MHz}$ frequency difference between their CLSs and the frequency reference light. Due to the bandwidth limitation of the piezo actuator of the laser for frequency locking, an acousto-optic modulator is inserted to increase the feedback bandwidth of the loop. In this way, two CLSs of Bob and Charlie have the same frequency, but are  $\SI{50}{\MHz}$ different from the Alice's MCWL. Thus, Alice needs to frequency shift her coherent light by  $\SI{50}{\MHz}$ with an acousto-optic modulator, so that the frequency of her CLS is the same as that of Bob's and Charlie's. Note that the MCWL is placed at Alice's end to simplify the demonstration. For geographically separated parties, it can instead be deployed at the measurement site, with the generated frequency reference beams transmitted to each party via additional optical fibers to enable frequency locking.

The encoders are the same for all three parties and are used to achieve weak coherent pulse (WCP) generation, pulse intensity modulation and phase modulation. The generation of coherent light pulses is realized by an intensity modulator, which converts continuous wave into light pulses of  $\SI{625}{\MHz}$ with a pulse width of about  $\SI{0.4}{\ns}$. The coherent light pulses are modulated into phase reference pulses and signal pulses by two Sagnac rings (SR) and a variable optical attenuator. The SR is a ring in which a phase modulator and a beam splitter are connected, and the phase modulator is in an asymmetric position to modulate different phases for the clockwise and counterclockwise pulses, thereby achieving the intensity modulation. The circulator in front of each SR is used to prevent the optical pulse from returning to the laser. One port of the circulator remains as the test port. The variable optical attenuator attenuates the light pulse to the single-photon level, assisting two SRs in modulating the required pulses of different intensities. The prepared signal pulses with a repetition rate of  $\SI{400}{\MHz}$ have three intensities corresponding to the signal state and two decoy states. The phase modulator between the two SRs is used for random phase modulation and phase coding. The signal pulses of the three parties are modulated with 16 different phases. As for the phase reference pulses, Bob and Charlie are required to modulate the pulse with three phases: 0, $\pi/2$ and $-\pi/2$, while Alice does not require phase modulation.

When the WCPs prepared by three parties arrive at the measurement site through three single-mode fiber channels, Alice's WCPs are first split into two beams to interfere with  WCPs of Bob and Charlie, forming two measurement branches. The WCPs entering the two measurement branches are aligned with polarization through the polarization feedback systems, and then enter the beam splitter for interference. Finally, the interference results are detected by the four channels of superconducting nanowire single-photon detector.  

Due to the imperfect frequency locking and disturbance of the fiber channel, the initial phase difference $\Delta \theta_{AB}$ ($\Delta \theta_{AC}$) between the WCPs prepared by Alice and Bob (Charlie) drift over time. The phase reference pulses are prepared and measured to track the phase drift. The statistical counts $n_{i}^{\phi}$ of four detector channels can be obtained every $\SI{100}{\us}$, where $i=1,\cdots , 4$, $\phi=0,\pi/2, -\pi/2$ corresponding to the phase reference pulses modulated the phase of 0, $\pi/2$ or $-\pi/2$ prepared by Bob and Charlie. In this way, $\cos \Delta \theta_{AB}$, $\sin \Delta \theta_{AB}$, $\cos \Delta \theta_{AC}$, and $\sin \Delta \theta_{AC}$ can be calculated and then $\Delta \theta_{AB}$, $\Delta \theta_{AC}$ can be obtained for phase compensation every $\SI{100}{\us}$. 

\begin{figure}[t]
	\includegraphics[width=\linewidth]{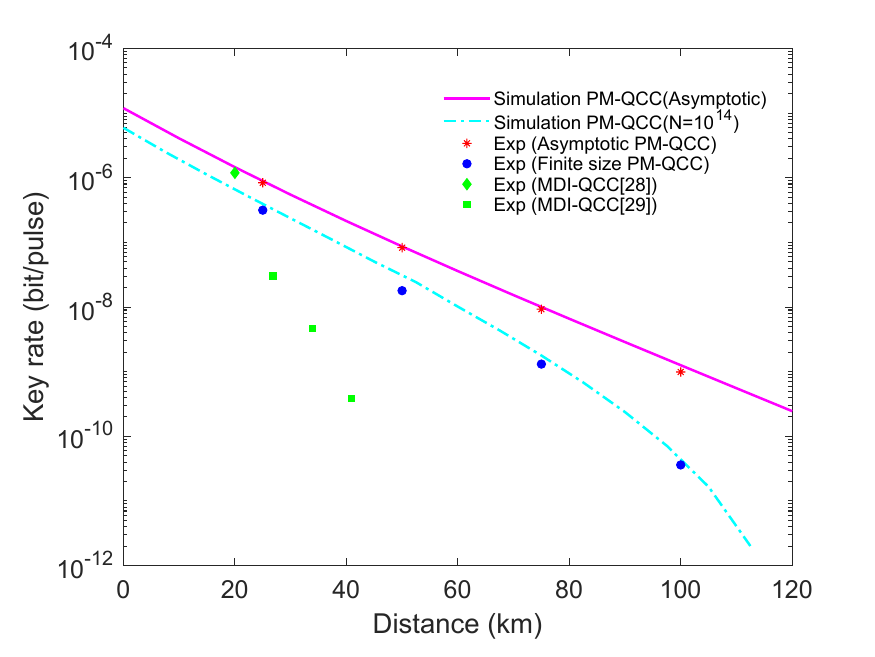}
	\caption{Key rates for symmetric fiber channels. The magenta solid line and cyan dash-dotted line represent the simulated asymptotic key rates and the key rates with total rounds of $N=10^{14}$ considering finite-size effects, respectively. The misalignment error $e_d=0.03$. The red asterisks and blue solid circles represent the experimental asymptotic key rates and key rates considering finite-size effects, respectively. The green diamonds and squares represent the experimental results of Refs. \cite{PRLYang2024} and \cite{PRLDu2025}, respectively. For comparison, attenuation used in the experiment of Ref. \cite{PRLDu2025} are converted to fiber distances with the same attenuation. The error correction efficiency $f = 1.06$ \cite{QIPTang2021}. The failure probability of finite-size analysis $\epsilon=10^{-10}$.}
\label{fig:symkeyrate}
\end{figure}

\section{Results}
In the demonstration of the three-intensity PM QCC protocol, the distances for three fiber channels are represented by $\{d_A, d_B, d_C\}$, where $d_A$, $d_B$, and $d_C$ are the distances from Alice, Bob, and Charlie to the measurement site, respectively. For the symmetric fiber channels, the key rates with total rounds of $N=10^{14}$ considering the finite-size effects are simulated, as shown in Fig.~\ref{fig:symkeyrate}. In the simulation, due to the splitting of Alice's pulse at the measurement site, the intensity of the signal state and two decoy states prepared by Alice is twice that of Bob and Charlie. The simulation results show that the maximum secure transmission distance from each party to the measurement site is over $\SI{100}{\km}$. Therefore, the three-intensity PM QCC protocol is demonstrated in the fiber configurations of $\{25, 25, 25\}$ km, $\{50, 50, 50\}$ km, $\{75, 75, 75\}$ km, and $\{100, 100, 100\}$ km, which is compatible with the intercity cryptographic applications. As shown in Table~\ref{table:exppr}, the total number of rounds corresponding to different fiber configurations are $2.14\times10^{13}$, $3.67\times10^{13}$, $4.66\times10^{13}$, and $1.28\times10^{14}$, and the maximum quantum bit error rates $E^\text{max}_Z$ are 3.26\%, 3.30\%, 3.27\%, and 3.59\%, respectively. The key rates considering finite-size effects and asymptotic key rates of the experiment is presented in Fig.~\ref{fig:symkeyrate}. It can be observed that the gap between the key rates considering finite-size effects and the asymptotic key rate increases with distance, even though $N$ for the channel configuration of $\{100, 100, 100\}$ km is over $10^{14}$. This indicates that the implementation of PM QCC protocol is severely limited by $N$ when the distance exceeds $\SI{100}{\km}$ due to finite-size effects. More advanced and efficient protocols are needed for further enhancements \cite{PRACarrara2023,CPXie2024,RoPiPLu2025}. Compared with the key rates of Refs.~\cite{PRLYang2024} and \cite{PRLDu2025}, our results have a significant improvement in the secure transmission distance and the key rate considering finite-size effects.

\begin{figure}[t]
  \includegraphics[width=\linewidth]{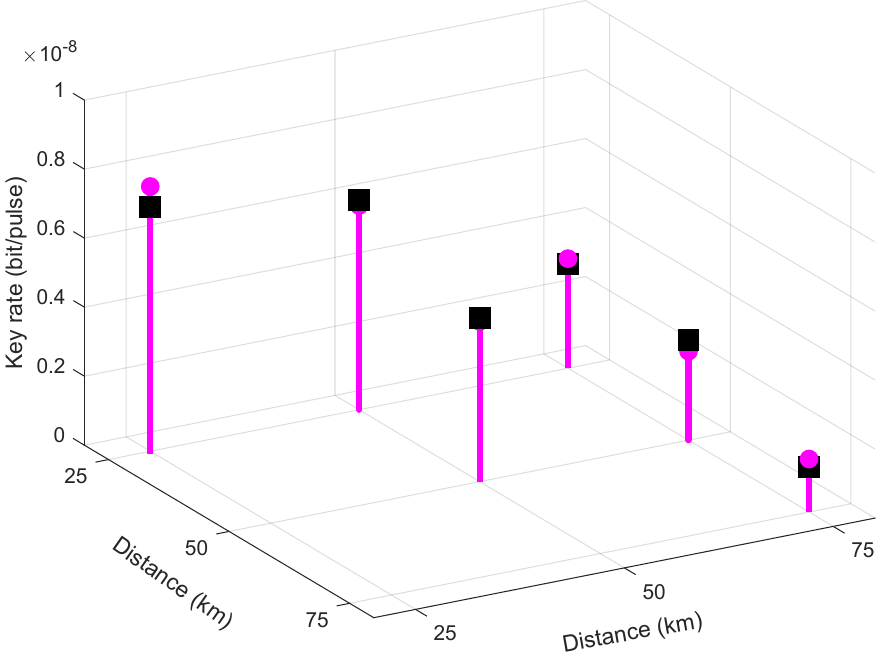}
  \caption{Key rates for asymmetric fiber channels. The black solid squares and magenta stems represent the experimental and simulated key rates, respectively. The total number of rounds $N=4\times10^{13}$ and the misalignment error $e_d=0.03$ for simulation. The secure key rate for symmetric fiber configurations of $\{75, 75, 75\}$ km are added for comparison.}
\label{fig:asymkr}
\end{figure}

For the asymmetric fiber channels, the three-intensity PM QCC protocol is demonstrated in the fiber configurations of $\{75, 25, 25\}$ km, $\{75, 50, 50\}$ km, $\{75, 25, 50\}$ km, $\{75, 25, 75\}$ km,  and $\{75, 50, 75\}$ km. The intensities of the signal state and the two decoy states of each party are optimized according to the channel loss while ensuring the same intensity ratio between the signal state and two decoy states. Similar to the fiber configuration of $\{75, 75, 75\}$ km, the total number of rounds of these fiber configurations is over $4\times10^{13}$. The key rates considering finite-size effects are shown in Fig. \ref{fig:asymkr}, where asymmetric fiber configurations show higher key rates compared to symmetric channels, with the $\{75,25,25\}$ km fiber configuration achieving a 5-fold increase over the $\{75,75,75\}$ km fiber configuration, eliminating the need for loss compensation in asymmetric channels. These results verify the feasibility of the three-intensity PM QCC protocol applied in real asymmetric networks. 

\begin{table*}[bpth!]
	\centering
	\caption{ Experimental parameters and results for symmetric and asymmetric channels. The detector dark count rate $p_d$ is about $2.4\times10^{-8}/\text{pulse}$. The total detection efficiency $\eta_d$ considering the insertion loss, the efficiency of SNSPD, the non-overlapping of the coherent pulse with the detection time window is about $60.5\%$. $L$ is the length of fiber channel configuration $\{d_A, d_B, d_C\}$. $R_1$ is the key rate considering the finite-size effects.}
     \scalebox{0.95}{
	\begin{tabular}{c|cccc|ccccc}
	\hline
	\hline
$L$ & $\{25,25,25\}$ & $\{50,50,50\}$ & $\{75,75,75\}$ & $\{100,100,100\}$ & $\{75,50,50\}$ & $\{75,25,25\}$ & $\{75,50,75\}$ & $\{75,25,75\}$ & $\{75,25,50\}$ \\
	\hline   
	$E_Z^{AB}$ & 3.26\% & 3.15\% & 3.27\% & 3.59\% & 3.27\% & 3.22\% & 3.16\% & 3.18\% & 3.31\% \\
	$E_Z^{AC}$ & 3.20\% & 3.30\% & 3.26\% & 3.45\% & 3.35\% & 3.27\% & 3.15\% & 3.20\% & 3.32\% \\
	$E_X^U$ & 13.43\%& 18.09\%&  18.12\%& 17.37\%& 17.14\% & 17.66\%& 17.01\% & 17.83\% & 17.01\%\\	
$N$ &  $2.14\times10^{13}$  &  $3.67\times10^{13}$  &  $4.66\times10^{13}$  &  $1.28\times10^{14}$  &  $4.39\times10^{13}$  &  $4.25\times10^{13}$  &  $4.58\times10^{13}$  &  $4.32\times10^{13}$  &  $4.18\times10^{13}$  \\
$R_1$ & $3.15\times10^{-7}$ & $1.79\times10^{-8}$ & $1.32\times10^{-9}$ & $4.20\times10^{-11}$ & $4.73\times10^{-9}$ & $6.83\times10^{-9}$ & $2.90\times10^{-9}$ & $3.00\times10^{-9}$ & $6.07\times10^{-9}$\\
	\hline    
	\hline    	
	\end{tabular}}
\label{table:exppr}
\end{table*}

\section{Conclusion}
The refined three-intensity PM QCC protocol for three parties has been applied to adapt to both symmetric and asymmetric fiber channels considering finite-size effects for experimental demonstration. In the symmetric fiber channels, the maximum secure transmission distance from each party to the measurement site reaches $\SI{100}{\km}$. In other words, the maximum fiber distance between the two parties is up to $\SI{200}{\km}$, which extends the secure transmission distance of the QCC protocol from the metropolitan area to the intercity distance. In the asymmetric fiber channels with a maximum channel distance of $\SI{75}{\km}$, higher secure key rates can be achieved using the optimized light intensities compared to the symmetric fiber channels with each channel distance of $\SI{75}{\km}$. This result verifies the feasibility of implementing PM QCC protocols for asymmetric channels in real networks without loss compensation. As our results provide a solid validation on the feasibility of implementing QCC in a real intercity star-type quantum network, we expect that this work can play a role in enriching and extending the functions and applications of current quantum networks.  

\begin{acknowledgments}
\section{Acknowledgments} 
This work was supported by the Quantum Science and Technology-National Science and Technology Major Project (Grant No.~2021ZD0300702), the Anhui Initiative in Quantum Information Technologies. S. Z. acknowledges support from the Zhejiang Provincial Natural Science Foundation of China (Grant No.~LQ24A050005) and the Quantum Science and Technology-National Science and Technology Major Project (Grant No.~2024ZD0302200). Y. M. acknowledges support from the National Natural Science Foundation of China (Grant No.~12104444) and the China Postdoctoral Science Foundation (Grant No.~2021M693093).

M. Z. and B.-C. L. contributed equally to this work.
\end{acknowledgments}

\appendix

\section{PM QCC Protocol and Security Analysis} \label{sec:sa}
In this section, we first present the steps of the three-intensity PM QCC protocol. Then, we present the security foundations and prove the security of the protocol using two different methods: the entanglement-based method and the source replacement method. Next, we develop a refined decoy state method that reduces the required number of intensities for the coherent state prepared by each party, and combine the finite key analysis to estimate the upper bound of the phase error rate. Finally, we simulate the performance of the protocol.
\subsection{Description of the PM QCC protocol}

Below, we present the steps of the three-intensity PM QCC protocol for three parties.

\begin{figure}[htbp]
	\includegraphics[width=0.9\linewidth]{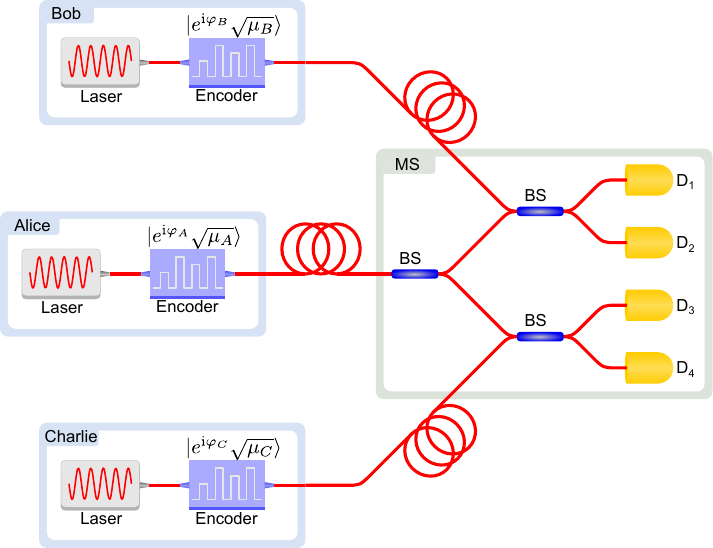}
	\caption{PM QCC protocol for three parties. Three parties Alice, Bob, and Charlie have the same device to prepare weak coherent state. They use lasers as coherent light sources and encoders to modulate the intensity and phase of coherent light pulses, thereby preparing the required coherent states $\ket{e^{\i\varphi_A}\sqrt{\mu_A}}$,$\ket{e^{\i\varphi_B}\sqrt{\mu_B}}$ and $\ket{e^{\i\varphi_C}\sqrt{\mu_C}}$. The coherent states prepared by the three parties are transmitted through fiber channels to the measurement site (MS). There, the coherent states prepared by Alice are split into two parts by beam splitter (BS), which interfere with the coherent states of Bob and Charlie, respectively, forming two measurement branches. The interference results of two branches are detected by four single-photon detectors (SPDs).}
\label{fig:protocol}
\end{figure}

\begin{enumerate}
	\item[Step.1] \textbf{Preparation}: As shown in Fig.~\ref{fig:protocol}, Alice encodes the laser pulse into the coherent state $\ket{e^{\i\varphi_A}\sqrt{\mu_A}}$ with an intensity $\mu_A$ randomly chosen from $\{ \mu_a , \nu_a ,\omega_a \}$ with probabilities $\{ p_\mu, p_\nu, p_\omega\}$, where $\mu_a$ is the intensity of signal state, $\nu_a$ and $\omega_a$ are intensities of two decoy states, $\omega_a<\nu_a<\mu_a$,  and $p_\mu+ p_\nu+p_\omega=1$, and a random phase $\varphi_A=\phi_{a}+\pi k_{a}$, where $k_{a}\in\{0,1\}$ is one random bit, $\phi_a$ is randomly chosen from $ \{ 0,\frac{2\pi}{D},\frac{4\pi}{D}\cdots,\frac{2\pi(D-1)}{D} \}$. Here, $D$ is the number of phase slices. Similarly, Bob (Charlie) prepares coherent state $\ket{e^{\i\varphi_B}\sqrt{\mu_B}}$ ($\ket{e^{\i\varphi_C}\sqrt{\mu_C}}$) with the intensity $\mu_B \in \{\mu_b,\nu_b,\omega_b\}$ ($\mu_C\in \{\mu_c,\nu_c,\omega_c\}$) and the phase $\varphi_B=\phi_b+\pi k_b$ ($\varphi_C=\phi_c+\pi k_c$). Here, $k_b$ ($k_c$) is a random bit similar to $k_a$, and $\phi_b$ ($\phi_c$) is a random phase similar to  $\phi_a$.

	\item[Step.2] \textbf{Measurement}:  All parties send their coherent states to the measurement site, an untrusted party Eve. An honest Eve splits each coherent pulse from Alice into two coherent pulses using a $50:50$ beam splitter to interfere with coherent pulses from Bob and Charlie using two $50:50$ beam splitters. The interference results  of the two measurement branches formed are detected by four single-photon detectors. Eve records successful detection events. A successful detection event is defined here as a coincident click of two measurement branches, where only one detector clicks for each measurement branch.
	
	\item[Step.3] \textbf{Announcement}: Eve announces all the successful detection events, and three parties announce their random phases $\phi_a$, $\phi_b$, $\phi_c$ and intensities $\mu_A$, $\mu_B$, $\mu_C$  of the coherent state corresponding to each event, respectively.
	
	\item[Step.4] \textbf{Sifting}: According to announcements, Alice, Bob, and Charlie keep their random bits only when the phase-matching conditions are satisfied: $|\phi_a-\phi_b|=0~\text{or}~\pi$, $|\phi_a-\phi_c|=0~\text{or}~\pi$, and their intensities are $\{ \mu_a,\mu_b,\mu_c \}$, $ \{ \nu_a,\nu_b,\nu_c \}$ or $\{ \omega_a,\omega_b,\omega_c \}$. Then, the random bits kept by Bob and Charlie remain unchanged or are flipped, depending on the type of detection event and the announced phase result.
	
	\item[Step.5] \textbf{Parameter estimation and key distillation}: The above steps are repeated enough times to distill the raw key bits. The quantum bit error rates (QBER) $E^{AB}_Z$ and $E^{AC}_Z$ and the gains $Q_{\mu}$, $Q_{\nu}$, $Q_{\omega}$ can be obtained directly from the experimental results. When considering the finite-size effects, the upper and lower bounds of gains $Q_{\mu}$, $Q_{\nu}$, $Q_{\omega}$ can be estimated by the Chernoff-Hoeffding method \cite{PRAZhang2017}. The upper bound of the phase error rate $E^U_X$  can be estimated using decoy-state method. Finally, Three parties distill private key bits by performing error correction and privacy amplification on the raw key.
\end{enumerate}

\subsection{Distillation of the GHZ state} \label{sec:GHZ}

The security of multiparty quantum information protocol is contingent upon the distribution of multiparty entangled states. The PM QCC protocol is equivalent to distributing Greenberger-Horne-Zeilinger (GHZ) state
\begin{equation}
\ket{\Phi^+} = \frac{1}{\sqrt{2}} \left( \ket{00\cdots 0} + \ket{11\cdots 1} \right)
\end{equation}
directly to the participants. The secure key rate of the protocol can be obtained from the distillation efficiency of the GHZ state.

The corresponding $M$-qubit GHZ state basis is
\begin{equation}
\ket{\Psi} = \frac{1}{\sqrt{2}} \left( \ket{0 i_1 \cdots i_{M-1}} + (-1)^j \ket{1 \bar{i}_1\cdots \bar{i}_{M-1}} \right),
\end{equation}
where $j (i_m) \in \{0,1\}$ and $\bar{i}_m$ is the logical negation of $i_m$. If $j=1$ ($i_m = 1$), there is a phase (bit) error to the original GHZ state. With use of the multipartite-hashing method, the yield of distillation of the pure $M$-qubit GHZ state is
\begin{equation}
Y = 1 - \max[h(E^{1,2}_Z),h(E^{1,3}_Z),\cdots,h(E^{1,M}_Z)] - h(E_X),
\end{equation}
where $E^{1,m}_Z$ represents the bit-flip error rate of the first part and $m$-th part, $E_X$ is the phase-flip error and $h(x) = -x \log_2(x) - (1-x) \log_2(1-x)$ is the binary entropy function.

\subsection{Entanglement-based method}\label{sec:ebm}
\begin{figure*}[htbp]
	\includegraphics[width=0.6\linewidth]{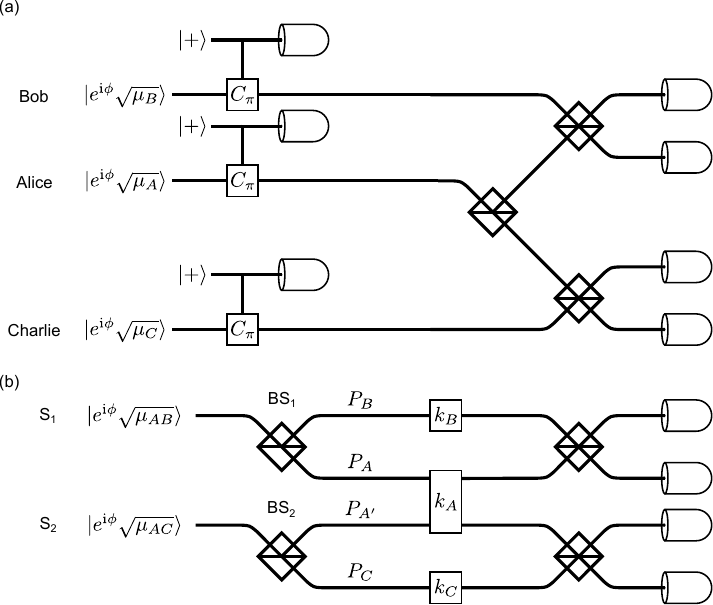}
	\caption{(a) The entanglement-based PM QCC protocol for three parties. (b) The equivalent entanglement-based PM QCC protocol for three parties with virtual sources after beam splitting.}
	\label{fig:ebm}
\end{figure*}
As shown in Fig. \ref{fig:ebm}, without loss of generality, an entanglement-based PM QCC protocol for three parties that is equivalent to the prepare-and-measure PM QCC protocol is presented \cite{PRAZhao2020}. There is a virtual qubit at each party
\begin{equation} \label{eq:virtual_qubit}
\ket{+} = \frac{1}{\sqrt{2}}\left( \ket{0} + \ket{1} \right)
\end{equation}
in Fig. \ref{fig:ebm} (a). Each party prepares an entanglement state using a controlled phase gate $C_\pi$ 
\begin{equation} \label{eq:controlled_phase_gate}
C_\pi = \ket{0}\bra{0}U_0 + \ket{1}\bra{1}U_\pi
\end{equation}
between the virtual qubit and the coherent state, where $U_{0(\pi)}$ will attach a phase of $0(\pi)$ to the coherent state. The entanglement state prepared by the $m$-th participant is
\begin{equation} \label{eq:WCP}
\ket{\Psi}_m = \frac{1}{\sqrt{2}} \left( \ket{0}\ket{e^{\i \phi}\sqrt{\mu_m}} + \ket{1}\ket{e^{\i (\phi+\pi)}\sqrt{\mu_m}} \right),
\end{equation}
where the coherent state is sent to an untrusted party to perform interference measurement with the coherent state prepared by other parties, while the virtual qubit is kept at each party.
\begin{widetext}

This is equivalent to another entanglement-based version shown in Fig. \ref{fig:ebm} (b), where the initial quantum state is written as
\begin{equation}
\ket{+}_{A}\ket{+}_{B}\ket{+}_{C}\sum_{n_1=0}^{\infty}e^{-\mu_{AB}/2}\frac{\left(e^{\i \phi}\sqrt{\mu_{AB}}C^\dagger_1\right)^{n_1}}{n_1!}\sum_{n_2=0}^{\infty}e^{-\mu_{AC}/2}\frac{\left(e^{\i \phi}\sqrt{\mu_{AC}}C^\dagger_2\right)^{n_2}}{n_2!} \ket{\text{vac}},
\end{equation}
where  $C^\dagger_i$ is the creation operator of the $i$th virtual source, $\ket{\text{vac}}$ is the vacuum state, $\mu_{AB} = \frac{\mu_A}{2} + \mu_B$ and $\mu_{AC} = \frac{\mu_A}{2} + \mu_C$. The evolution matrix of BS\textsubscript{1} and BS\textsubscript{2} can be expressed as
\begin{equation} \label{eq:BS}
\begin{split}
T_\text{BS\textsubscript{1}} &= 
\begin{bmatrix}
	r_1 & t_1 \\
	t_1 & -r_1
	\end{bmatrix},\\
T_\text{BS\textsubscript{2}} &= 
\begin{bmatrix}
	r_2 & t_2 \\
	t_2 & -r_2
	\end{bmatrix},
\end{split}
\end{equation}
where $r_1 = \sqrt{\frac{\mu_B}{\mu_A/2+\mu_B}}$, $t_1 = \sqrt{\frac{\mu_A/2}{\mu_A/2+\mu_B}}$, $r_2 = \sqrt{\frac{\mu_A/2}{\mu_A/2+\mu_C}}$, and $t_2 = \sqrt{\frac{\mu_C}{\mu_A/2+\mu_C}}$. The quantum state takes the following form
\begin{equation}
\sum_{n_1,n_2=0}^{\infty} \ket{+}_{A}\ket{+}_{B}\ket{+}_{C} \frac{e^{-\mu_{AB}/2}(\mu_{AB})^{n_1/2}}{n_1!} \left( r_1 P^\dagger_B e^{\i \phi} + t_1 P^\dagger_{A} e^{\i \phi} \right)^{n_1} \frac{e^{-\mu_{AC}/2}(\mu_{AC})^{n_2/2}}{n_2!} \left( r_2 P^\dagger_{A'} e^{\i \phi} + t_2 P^\dagger_{C} e^{\i \phi} \right)^{n_2} \ket{\text{vac}}. 
\end{equation}
after passing through two beam splitters BS\textsubscript{1} and BS\textsubscript{2}. Here $P^\dagger_{A}$ ($P^\dagger_{A'}$) and $P^\dagger_{B}$ ($P^\dagger_{C}$) are the creation operators of the corresponding path mode after beam splitters BS\textsubscript{1} (BS\textsubscript{2}). 
A $C_\pi$ gate is used for bit encoding, then the quantum state becomes
\begin{equation}
\begin{split}
\frac{1}{2^{3/2}}&
\sum_{n_1,n_2=0}^{\infty} \frac{e^{-\mu_\text{tot}/2}(\mu_{AB})^{n_1/2}(\mu_{AC})^{n_2/2}}{n_1!n_2!} \left[ \ket{0}\ket{0}\ket{0} \left( r_1 P^\dagger_B e^{\i \phi} + t_1 P^\dagger_{A} e^{\i \phi} \right)^{n_1} \left( r_2 P^\dagger_{A'} e^{\i \phi} + t_2 P^\dagger_C e^{\i \phi} \right)^{n_2} + \cdots+ \right.\\
&\left.\ket{1}\ket{1}\ket{1}
\left( -r_1 P^\dagger_B e^{\i \phi} - t_1 P^\dagger_{A} e^{\i \phi} \right)^{n_1}\left( - r_2 P^\dagger_{A'} e^{\i \phi} - t_2 P^\dagger_C e^{\i \phi} \right)^{n_2} \right] \ket{\text{vac}},
\end{split}
\end{equation}
where $\mu_\text{tot} = \mu_A + \mu_B + \mu_C$. This quantum state is equivalent to the following quantum state

\begin{equation}
\begin{split}
&\frac{1}{2} \sum_{n_1,n_2=0}^{\infty} \frac{e^{-\mu_\text{tot}/2}(\mu_{AB})^{n_1/2}(\mu_{AC})^{n_2/2}}{n_1!n_2!}
\left\{ \sum_{i_A,i_C \in \{0,1\}} \frac{1}{\sqrt{2}}  \left[ \ket{0 i_A i_C} + (-1)^{n_1+n_2}\ket{1 \bar{i}_A \bar{i}_C} \right] \right.\\
&\left.\left[ r_1 P^\dagger_B e^{\i \phi} + (-1)^{i_A} t_1 P^\dagger_{A} e^{\i \phi} \right]^{n_1}
\left[ (-1)^{i_A} r_2 P^\dagger_{A'} e^{\i \phi} + (-1)^{i_C} t_2 P^\dagger_C e^{\i \phi} \right]^{n_2}  \right\}
\ket{\text{vac}}\\
&= \frac{1}{2} \sum_{n_1,n_2=0}^{\infty} \frac{e^{-\mu_\text{tot}/2}(\mu_{AB})^{n_1/2}(\mu_{AB})^{n2/2}}{n_1!n_2!}
\left\{ \sum_{i_A,i_C \in \{0,1\}} \frac{1}{\sqrt{2}}  \left[ \ket{0 i_A i_C} + (-1)^{n_1+n_2}\ket{1 \bar{i}_A \bar{i}_C} \right] \right.\\
&\left.\left[ r_1 P^\dagger_B e^{\i \phi} + (-1)^{i_A} t_1 P^\dagger_{A} e^{\i \phi} \right]^{n_1} 
\left[ (-1)^{i_A} r_2 P^\dagger_{A'} e^{\i \phi} + (-1)^{i_C} t_2 P^\dagger_C e^{\i \phi} \right]^{n_2}  \right\}
\ket{\text{vac}}\\
& = \frac{1}{2\sqrt{2}}\sum_{i_A,i_C\in \{0,1\}}\left\{\sum_{n_1,n_2 \in \text{even}} \frac{e^{-\mu_\text{tot}/2}(\mu_{AB})^{n_1/2}(\mu_{AC})^{n_2/2}}{n_1!n_2!} \left[ \ket{0i_A i_C}+\ket{1\bar{i}_A \bar{i}_C} \right] \right. \\
&\left.\left[ r_1 P^\dagger_B e^{\i \phi} + (-1)^{i_A} t_1 P^\dagger_{A} e^{\i \phi} \right]^{n_1} 
\left[ (-1)^{i_A} r_2 P^\dagger_{A'} e^{\i \phi} + (-1)^{i_C} t_2 P^\dagger_C e^{\i \phi} \right]^{n_2} +\right.\\
&   
\sum_{n_1,n_2 \in \text{odd}} \frac{e^{-\mu_\text{tot}/2}(\mu_{AB})^{n_1/2}(\mu_{AC})^{n_2/2}}{n_1!n_2!} \left[ \ket{0i_A i_C}+\ket{1\bar{i}_A \bar{i}_C} \right]\\
&\left.\left[ r_1 P^\dagger_B e^{\i \phi} + (-1)^{i_A} t_1 P^\dagger_{A} e^{\i \phi} \right]^{n_1} 
\left[ (-1)^{i_A} r_2 P^\dagger_{A'} e^{\i \phi} + (-1)^{i_C} t_2 P^\dagger_C e^{\i \phi} \right]^{n_2}
\right\}
\ket{\text{vac}}.\\
\end{split}
\end{equation}

The quantum states held by the three participating parties are ultimately determined to be
\begin{equation}
\frac{1}{2} \sum_{i_A,i_C \in \{0,1\}} \left\{ \frac{1}{\sqrt{2}} \left[ \ket{0 i_A i_C}
+ \ket{1 \bar{i}_A \bar{i}_C} \right] \sqrt{p_\text{even}} \ket{\text{even}} + \frac{1}{\sqrt{2}} \left[ \ket{0 i_A i_C} - \ket{1 \bar{i}_A \bar{i}_C} \right] \sqrt{p_\text{odd}} \ket{\text{odd}} \right\},
\end{equation}
where $\ket{\text{even}}$ and $\ket{\text{odd}}$ represent those states containing odd and even numbers of photons, respectively, which are given by
\begin{equation}
\begin{split}
\ket{\text{even}} = &\frac{1}{\sqrt{p_{\text{even}}}} \sum_{n_1,n_2 \in \text{even}} \frac{e^{-\mu_\text{tot}/2}(\mu_{AB})^{n_1/2}(\mu_{AC})^{n_2/2}}{n_1!n_2!}\\
&\left[ r_1 P^\dagger_B e^{\i \phi}+ (-1)^{i_A} t_1 P^\dagger_{A} e^{\i \phi} \right]^{n_1} \left[ (-1)^{i_A} r_2 P^\dagger_{A'} e^{\i \phi} + (-1)^{i_C} t_2 P^\dagger_C e^{\i \phi} \right]^{n_2} \ket{\text{vac}},\\
\ket{\text{odd}} = &\frac{1}{\sqrt{p_{\text{odd}}}} \sum_{n_1,n_2 \in \text{odd}} \frac{e^{-\mu_\text{tot}/2}(\mu_{AB})^{n_1/2}(\mu_{AC})^{n_2/2}}{n_1!n_2!}\\
&\left[ r_1 P^\dagger_B e^{\i \phi}+ (-1)^{i_A} t_1 P^\dagger_{A} e^{\i \phi} \right]^{n_1} \left[ (-1)^{i_A} r_2 P^\dagger_{A'} e^{\i \phi} + (-1)^{i_C} t_2 P^\dagger_C e^{\i \phi} \right]^{n_2} \ket{\text{vac}},\\
\end{split}
\end{equation}
and $\sqrt{p_\text{even}}$ and $\sqrt{p_\text{odd}}$ are normalized coefficients of pure states $\ket{\text{even}}$ and $\ket{\text{odd}}$, which are given by
\begin{equation}
\begin{split}
p_{\text{even}} &= \sum_{n_1,n_2 \in \text{even}} \frac{e^{-\mu_\text{tot}/2}(\mu_{AB})^{n_1/2}(\mu_{AC})^{n_2/2}}{n_1!n_2!} = e^{-\mu_\text{tot}} \cosh{(\mu_\text{tot})},\\
p_{\text{odd}} &= \sum_{n_1,n_2 \in \text{odd}} \frac{e^{-\mu_\text{tot}/2}(\mu_{AB})^{n_1/2}(\mu_{AC})^{n_2/2}}{n_1!n_2!} = e^{-\mu_\text{tot}} \sinh{(\mu_\text{tot})}.
\end{split}
\end{equation}
\end{widetext}

Thus, according to the description in Sec.~\ref{sec:GHZ}, the phase error rate $E_{X}$ can be estimated by
\begin{equation} \label{eq:phase_error}
E_{X} = p_\text{odd} \frac{Y_\text{odd}}{Q_\mu},
\end{equation}
where $Y_\text{odd}$ is the overall yield for odd-number components, and $Q_\mu$ is the overall gain of signal state.

\subsection{Source replacement method}\label{sec:srm}
\begin{figure*}[htbp]
	\includegraphics[width=0.6\linewidth]{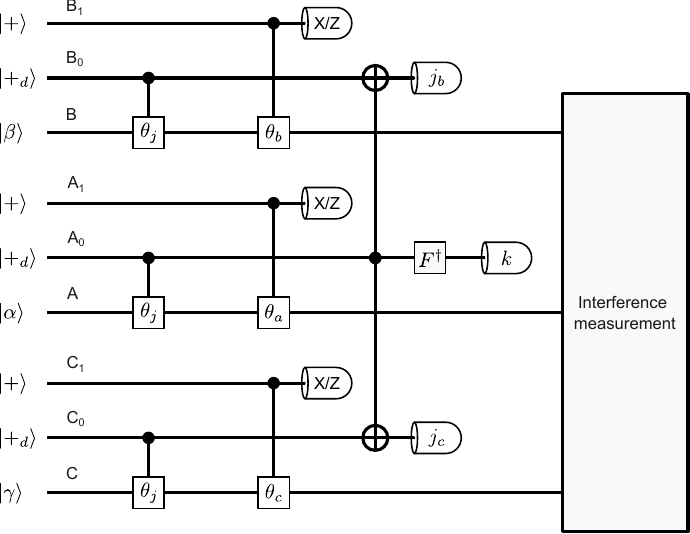}
	\caption{Encoding circuit of source-replaced scheme. The auxiliary systems $A_0$, $A_1$, $B_0$, $B_1$, $C_0$ and $C_1$ are used to encode the random phases and key bits. Subsequent to the encoding of coherent states, the auxiliary systems undergo gate operations and measurements, which can be utilized to determine the phase and bit error rates of the entangled system. The gate operations involved in this process include a quantum inverse Fourier transform, denoted by $F^\dagger$, applied to the auxiliary system of Alice, together with two high-dimensional CNOT operations.}
	\label{fig:srm}
\end{figure*}

As shown in Fig.~\ref{fig:srm},  we employ the source replacement method \cite{AQTHuang2024} to make the security proof of the PM QCC protocol more intuitive. A protocol equivalent to PM QCC can be represented as a quantum gate encoding circuit. The initial state of circuit can be expressed as
\begin{equation}
\ket{+_d}_{A_0}\ket{+}_{A_1}\ket{\alpha}_{A}\ket{+_d}_{B_0}\ket{+}_{B_1}\ket{\beta}_{B}\ket{+_d}_{C_0}\ket{+}_{C_1}\ket{\gamma}_{C}, 
\end{equation}
where $\ket{+} = \frac{1}{\sqrt{2}}(\ket{0}+\ket{1})$, $\ket{+_d}$ is the eigen state of Pauli X operator in d-dimension and $\ket{\alpha}$, $\ket{\beta}$,  and $\ket{\gamma}$ are coherent state. The Z-basis measurement results at $\ket{+}$ and $\ket{+_d}$ control the key bit encoding using $C-\theta_a$, $C-\theta_b$, and $C-\theta_c$ gates and the random phase encoding using $C-\theta_j$ gate, respectively. 
\begin{widetext}
Once the initial quantum state is operated by these C-phase gates, the auxiliary systems of Alice, Bob, and Charlie are entangled with the optical modes sent for interference measurement. The initial quantum state is transformed into

\begin{equation}\label{eq:pes}
\begin{split}
\left( \frac{1}{2d} \right)^{\frac{3}{2}}&\left[ \sum_{j_a = 0}^{d-1} \ket{j_a}_{A_0} \left( \ket{0}_{A_1}
\ket{e^{2\pi \i  j_a /d}\alpha}_A + \ket{1}_{A_1}\ket{-e^{2\pi \i  j_a /d}\alpha}_A \right) \right] \\
& \left[ \sum_{j_b = 0}^{d-1} \ket{j_b}_{B_0} \left( \ket{0}_{B_1}\ket{e^{2\pi \i  j_b /d}\beta}_B + \ket{1}_{B_1}\ket{-e^{2\pi \i  j_b /d}\beta}_B \right) \right]\\ 
&\left[ \sum_{j_c = 0}^{d-1} \ket{j_c}_{C_0} \left( \ket{0}_{C_1}\ket{e^{2\pi \i  j_c /d}\gamma}_C + \ket{1}_{C_1}\ket{-e^{2\pi \i  j_c /d}\gamma}_C \right) \right].
\end{split}
\end{equation}

Following the operation of two CNOT gates, the quantum state undergoes unitary evolution into

\begin{equation}
\begin{split}
&
\left( \frac{1}{2d} \right)^{\frac{3}{2}} 
\left[ \sum_{j_a = 0}^{d-1} \ket{j_a}_{A_0}
\left( \ket{0}_{A_1}\ket{e^{2\pi \i  j_a /d} \alpha }_A + \ket{1}_{A_1}\ket{-e^{2\pi \i  j_a /d}\alpha}_A \right) \right] \\
& 
\left[ \sum_{j_b = 0}^{d-1} \ket{j_b - j_a}_{B_0} \left( \ket{0}_{B_1}\ket{e^{2\pi \i  j_b /d} \beta }_B + \ket{1}_{B_1}\ket{-e^{2\pi \i  j_b /d}\beta}_B \right) \right] \\
&
\left[ \sum_{j_c = 0}^{d-1} \ket{j_c - j_a}_{C_0} \left( \ket{0}_{C_1}\ket{e^{2\pi \i  j_c /d} \gamma }_C + \ket{1}_{C_1}\ket{-e^{2\pi \i  j_c /d}\gamma}_C \right) \right] \\
&
= \left( \frac{1}{2d} \right)^{\frac{3}{2}} \sum_{j_1 = 0}^{d-1} \sum_{j_2 = 0}^{d-1} \sum_{j_a = 0}^{d-1}
\ket{j_a}_{A_0}
\left( \ket{0}_{A_1}\ket{e^{2\pi \i  j_a /d}\alpha}_A + \ket{1}_{A_1}\ket{-e^{2\pi \i  j_a /d}\alpha}_A \right) \\
& 
\ket{j_1}_{B_0}\left( \ket{0}_{B_1}\ket{e^{2\pi \i  (j_1+j_a) /d}\beta}_B + \ket{1}_{B_1}\ket{-e^{2\pi \i  (j_1+j_a) /d}\beta}_B \right)\\
&
\ket{j_2}_{C_0}\left( \ket{0}_{C_1}\ket{e^{2\pi \i  (j_2+j_a) /d}\gamma}_C + \ket{1}_{C_1}\ket{-e^{2\pi \i  (j_2+j_a) /d}\gamma}_C \right),
\end{split}
\end{equation}

where $j_1 = j_b - j_a$ and $j_2 = j_c - j_a$. Acting a quantum inverse Fourier transform gate $F^\dagger$ on $A_0$ results in

\begin{equation}
\begin{split}
&\left( \frac{1}{2} \right)^{\frac{3}{2}} \left( \frac{1}{d} \right)^2 \sum_{j_1=0}^{d-1}\sum_{j_2=0}^{d-1}\sum_{j_a=0}^{d-1}\sum_{k=0}^{d-1}e^{-2\pi \i  k j_a/d}\\
&\ket{k}_{A_0}
\left( \ket{0}_{A_1}\ket{-e^{2\pi \i  j_a /d}\alpha}_A + \ket{1}_{A_1}\ket{-e^{2\pi \i  j_a /d}\alpha} _A\right) \\
& \ket{j_1}_{B_0}\left( \ket{0}_{B_1}\ket{e^{2\pi \i  (j_1+j_a) /d}\beta}_B + \ket{1}_{B_1}\ket{-e^{2\pi \i  (j_1+j_a) /d}\beta}_B \right)\\
&\ket{j_2}_{C_0}\left( \ket{0}_{C_1}\ket{e^{2\pi \i  (j_2+j_a) /d}\gamma}_C + \ket{1}_{C_1}\ket{-e^{2\pi \i  (j_2+j_a) /d}\gamma}_C \right),
\end{split}
\end{equation}
where
\begin{equation}
F^\dagger = \frac{1}{\sqrt{d}}\sum_{j=0}^{d-1}\sum_{k=0}^{d-1}e^{-2\pi \i j k/d}\ket{k}\bra{j}.
\end{equation} 
Subsequent to the implementation of projective measurements on the auxiliary systems $A_0$, $B_0$, and $C_0$ in their respective basis, the final system evolves into

\begin{equation}
\begin{split}
&
\sum_{j_a = 0}^{d-1} e^{-2\pi \i kj_a/d}
\left( \ket{0}_{A_1} \ket{e^{2\pi \i  j_a /d}\alpha}_A + \ket{1}_{A_1} \ket{-e^{2\pi \i  j_a /d}\alpha}_A \right) \\
&
\left( \ket{0}_{B_1} \ket{e^{2\pi \i  j_a /d}\beta '}_B + \ket{1}_{B_1} \ket{-e^{2\pi \i  j_a /d}\beta '}_B \right)
\left( \ket{0}_{C_1} \ket{e^{2\pi \i  j_a /d}\gamma '}_C + \ket{1}_{C_1} \ket{-e^{2\pi \i  j_a /d}\gamma '}_C \right) \\
&
= e^{-(|\alpha|^2+|\beta|^2+|\gamma|^2)/2} \sum_{j_a=0}^{d-1} \sum_{n=0}^{\infty}\sum_{m=0}^{\infty}\sum_{l=0}^{\infty} e^{2\pi \i (n+m+l-k)j_a/d}\\
&
\left[\ket{0}+(-1)^n\ket{1}\right]_{A_1}
\left[\ket{0}+(-1)^m\ket{1}\right]_{B_1}
\left[\ket{0}+(-1)^l\ket{1}\right]_{C_1} \frac{\alpha^n\beta'^m\gamma'^l}{\sqrt{n!m!l!}}\ket{nml}_{ABC}.
\end{split}
\end{equation}
\end{widetext}
Here, $e^{2\pi\i j_b/d}\beta$ is denoted as $\beta'$ and $e^{2\pi\i j_c/d}\gamma$ is denoted as $\gamma'$ for clarity. After performing the interference measurement on coherent state, if $n+m+l$ is even (odd), the state shared by Alice, Bob and Charlie is the linear combination of $\frac{1}{\sqrt{2}}\left(\ket{0 i j} + \ket{1 \bar{i} \bar{j}}\right)$ ($\frac{1}{\sqrt{2}}\left(\ket{0 i j} - \ket{1 \bar{i} \bar{j}}\right)$). The key bits can be obtained by performing Z-basis measurements on the auxiliary systems $A_1$, $B_1$, and $C_1$, whose phase error rate can be obtained by performing X-basis measurements on these auxiliary systems. According to the analysis of Sec.~\ref{sec:GHZ}, the phase error rate expressed as Eq.~\eqref{eq:phase_error} can be obtained.

\subsection{Decoy-state method}\label{sec:dsm}
The phase errors rate $E_X$ in Eq.~\eqref{eq:phase_error} can be rewritten as
\begin{equation}
\begin{split}
E_X & = \sum_{k \in \text{odd}} P_\mu(k) \frac{Y_k}{Q_\mu} \\
& = 1- \sum_{k \in \text{even}} P_\mu(k) \frac{Y_k}{Q_\mu} \\
& \leq E_X^U = 1 - P_\mu(2) \frac{Y_2^L}{Q_\mu},
\end{split}
\end{equation}
where $P_\mu(k)$ is the probability that all parties emit $k$ photons, following a Poisson distribution with total intensity $\mu_\text{tot}$ as the parameter. $Y_2^L$ is the lower bound of the yield of two photons, which can be estimated by decoy-state method with the intensity set $\{ \mu , \nu , \omega , 0 \}$. Define
\begin{equation}
\begin{split}
\mu_\text{tot} &= \mu_a + \mu_b + \mu_c, \\
\nu_\text{tot} &= \nu_a + \mu_b + \nu_c, \\
\omega_\text{tot} &= \omega_a + \omega_b + \omega_c, \\
\mu_\text{tot} &> \nu_\text{tot} > \omega_\text{tot}.
\end{split}
\end{equation}
Here $\mu_a : \nu_a : \omega_a = \mu_b : \nu_b : \omega_b = \mu_c : \nu_c : \omega_c$ is assumed. Then
\begin{equation}
\begin{split}
e^{\mu_\text{tot}} Q_\mu &= Y_0 + 2\mu_\text{tot} Y_1 + \frac{\mu_\text{tot}^2}{2}Y_2 + \frac{\mu_\text{tot}^3}{6}Y_3 + \cdots, \\
e^{\nu_\text{tot}} Q_\nu &= Y_0 + 2\nu_\text{tot} Y_1 + \frac{\nu_\text{tot}^2}{2}Y_2 + \frac{\nu_\text{tot}^3}{6}Y_3 + \cdots, \\
e^{\omega_\text{tot}} Q_\omega &= Y_0 + 2\omega_\text{tot} Y_1 + \frac{\omega_\text{tot}^2}{2}Y_2 + \frac{\omega_\text{tot}^3}{6}Y_3 + \cdots, \\
Q_0 &= Y_0.
\end{split}
\end{equation}
Using the Gaussian-elimination method, we have
\begin{equation}
Y_2 \geq \frac{2}{G} \left( G_\nu Q_\nu + G_0 Q_0 - G_\mu Q_\mu - G_\omega Q_\omega \right),
\end{equation}
where 
\begin{equation}
\begin{split}
G =& (\mu_\text{tot} -  \nu_\text{tot})( \nu_\text{tot} - \omega_\text{tot}) \mu_\text{tot} \nu_\text{tot}^2 \omega_\text{tot}(\mu_\text{tot}-\omega_\text{tot})>0,\\
G_{\nu} =& \mu_\text{tot} \nu_\text{tot} \omega_\text{tot} e^{\nu_\text{tot}}(\mu_\text{tot}^2 - \omega_\text{tot}^2 )>0,\\
G_0  =&(\mu_\text{tot}-\nu_\text{tot}) (\nu_\text{tot}-\omega_\text{tot})\nu_\text{tot}\\
&\left[\mu_\text{tot} (\mu_\text{tot}+ \nu_\text{tot}) - \omega_\text{tot}(\nu_\text{tot}+ \omega_\text{tot})\right]>0,\\
G_\mu =& (\nu_\text{tot}^2 - \omega_\text{tot}^2)\nu_\text{tot}^2\omega_\text{tot}e^{\mu_\text{tot}}>0, \\
G_\omega =& (\mu_\text{tot}^2 - \nu_\text{tot}^2)\mu_\text{tot} \nu_\text{tot}^2e^{\omega_\text{tot}}>0.
\end{split}
\end{equation}
Thus $Q_0G_0\geq0$ and 
\begin{equation}
Y_2 \geq Y_2^L = \frac{2}{G} \left( G_\nu Q_\nu - G_\mu Q_\mu - G_\omega Q_\omega \right).
\end{equation}
Based on the yield of distillation of the pure 3-qubit GHZ state, the key rate of PM QCC protocol can be expressed as
\begin{equation} \label{eq:keyrate}
R = \left( \frac{2}{D} \right)^2 Q_\mu \left[1-fh(E_Z^{\max})-h(E^U_{X})\right],
\end{equation}
where $(2/D)^2$ is the prefactor induced by phase post-selection.

\subsection{Finite-size analysis}\label{sec:Finite_size_analysis}
Given an observed quantity $\chi$, follow the Chernoff-Hoeffding method in Ref.~\cite{PRAZhang2017}, the upper bound and lower bound of the underlying expectation value are given by
\begin{equation}
\begin{split}
\mathbb{E}^L(\chi)&=\dfrac{\chi}{1+\delta^L},\\
\mathbb{E}^U(\chi)&=\dfrac{\chi}{1-\delta^U},
\end{split}
\end{equation}
where $\delta^{L(U)}$ is the solution of
\begin{equation}\label{eq:deltasolve}
\begin{split}
\left[\dfrac{e^{\delta^L}}{(1+\delta^L)^{1+\delta^L}}\right]^{\chi / (1+\delta^L)}&=\dfrac{1}{2}\epsilon,\\
\left[\dfrac{e^{-\delta^U}}{(1-\delta^U)^{1-\delta^U}}\right]^{\chi / (1-\delta^U)}&=\dfrac{1}{2}\epsilon.
\end{split}
\end{equation}
The $\epsilon$ is failure probability of the estimation $\mathbb{E}(\chi)\in [\mathbb{E}^L(\chi), \mathbb{E}^U(\chi)]$. When finite-size effects are considered, the lower bound of the yield of two-photon state is given by
\begin{equation}\label{eq:finit_key_y2}
Y_2^L = \frac{2}{G} \left( G_\nu Q_\nu^L - G_\mu Q_\mu^U - G_\omega Q_\omega^U \right),
\end{equation}
where $Q_\mu^U$ and $Q_\omega^U$ are the upper bounds of $Q_\mu$ and $Q_\omega$, respectively, and $Q_\nu^L$ is the lower bound of $Q_\nu$.

\subsection{Simulation method and result}\label{sec:Simulation_method_and_result}
Firstly, we obtain the click probability in the PM QCC protocol, and then calculate the gain under the given information of the sender and the channel.  The intensity of coherent state prepared by Alice, Bob and Charlie are denoted as $\mu_A\in\{\mu_a, \nu_a, \omega_a\}$, $\mu_B\in\{\mu_b, \nu_b, \omega_b\}$ and $\mu_C\in\{\mu_c, \nu_c, \omega_c\}$, respectively. The transmittance from Alice, Bob, Charlie to the measuremnt site are denoted as $\eta_{A}$, $\eta_{B}$ and $\eta_{C}$.  The total detection efficiency of measuremnt site is $\eta_d$. The click probabilities of two measurement branches for each coherent pulse are given by
\begin{equation}
\begin{split}
	p_{\mu_A,\mu_B}^\text{click} &= 1- e^{-\eta_A\eta_d \mu_A/2 - \eta_B\eta_d  \mu_B} + \\
&2 p_d e^{-\eta_A\eta_d  \mu_A/2 - \eta_B\eta_d  \mu_B}, \\ 
	p_{\mu_A,\mu_C}^\text{click}&= 1- e^{-\eta_A\eta_d  \mu_A/2 - \eta_C\eta_d \mu_C} + \\
&2 p_d e^{-\eta_A\eta_d \mu_A/2 - \eta_C\eta_d  \mu_C},
\end{split}
\end{equation}
where $p_d$ represents the dark count rate of the single-photon detector. The gain can be written as
\begin{equation}
	Q_{\mu_A,\mu_B,\mu_C} = p_{\mu_A,\mu_B}^\text{click} p_{\mu_A,\mu_C}^\text{click}.
\end{equation}

The sending probabilities of signal state and two decoy states for each party are denoted as $p_{\mu}$, $p_{\nu}$, and $p_{\omega}$, respectively. The total number of rounds is denoted as $N$.  Thus, the number of rounds in which Alice, Bob, and Charlie simultaneously send signal and two decoy states are given by
\begin{equation}
	\begin{split}
		N_\mu &= N p_{\mu}^3, \\
		N_\nu &= N p_{\nu}^3, \\
		N_\omega &= N p_{\omega}^3.
	\end{split}
\end{equation}

The effective clicks for simultaneously sending signal and two decoy states are given by
\begin{equation}
	\begin{split}
		M_\mu &= N_{\mu} Q_{\mu_a,\mu_b,\mu_c} =  N_{\mu} Q_{\mu},\\
		M_\nu &= N_{\nu} Q_{\nu_a,\nu_b,\nu_c} = N_{\nu} Q_{\nu}, \\
		M_\omega &= N_{\omega} Q_{\omega_a,\omega_b,\omega_c} = N_{\omega} Q_{\omega}.
	\end{split}
\end{equation}

Based on the finite-key analysis method in Sec.~\ref{sec:Finite_size_analysis}, The statistical upper and lower bounds of the effective clicks can be calculated by the Chernoff bound method. Then the upper and lower bounds of the gain for each state can be obtained. By substituting it into Eq.~\eqref{eq:finit_key_y2}, the lower bound of the two-photon yield can be derived.

The upper bound of the phase error can be written as
\begin{equation}\label{eq:simulation_ex}
	E^U_X = 1 - e^{-\mu_\text{tot}} \frac{\mu_\text{tot}^2}{2 Y_2^L Q_\mu^L}.
\end{equation}

The quantum bit error rate can be expressed by  
\begin{equation}\label{eq:simulation_Ez}
	E_Z^\text{max} = \max\{e^{AB}, e^{AC}\} + e_d,
\end{equation}
where $e_d$ is the misalignment error, $e^{AB}$ and $e^{AC}$ are intrinsic errors of two measurement branch caused by dark counts and the discretization of the phase slice, which are given by
\begin{equation}
	\begin{split}
	e^{AB} = &\frac{1}{p_{\mu_a,\mu_b}^\text{click}} \left[p_d + (\eta_{A}\eta_d \mu_a/2 + \eta_{B}\eta_d \mu_b)e_\delta \right]\\
& e^{-\eta_{A}\eta_d \mu_a/2 - \eta_{B}\eta_d \mu_b},\\
	e^{AC} = &\frac{1}{p_{\mu_a,\mu_c}^\text{click}} \left[p_d + (\eta_{A}\eta_d \mu_a/2 + \eta_{C}\eta_d \mu_c)e_\delta \right] \\
&e^{-\eta_{A}\eta_d \mu_a/2 - \eta_{C}\eta_d \mu_c},\\
	e_\delta = &\frac{\pi}{D} - \left(\frac{D}{\pi}\right)^2 \left( \sin \frac{\pi}{D} \right)^3.
	\end{split}
\end{equation}

\begin{figure*}[htbp]
	\includegraphics[width=\linewidth]{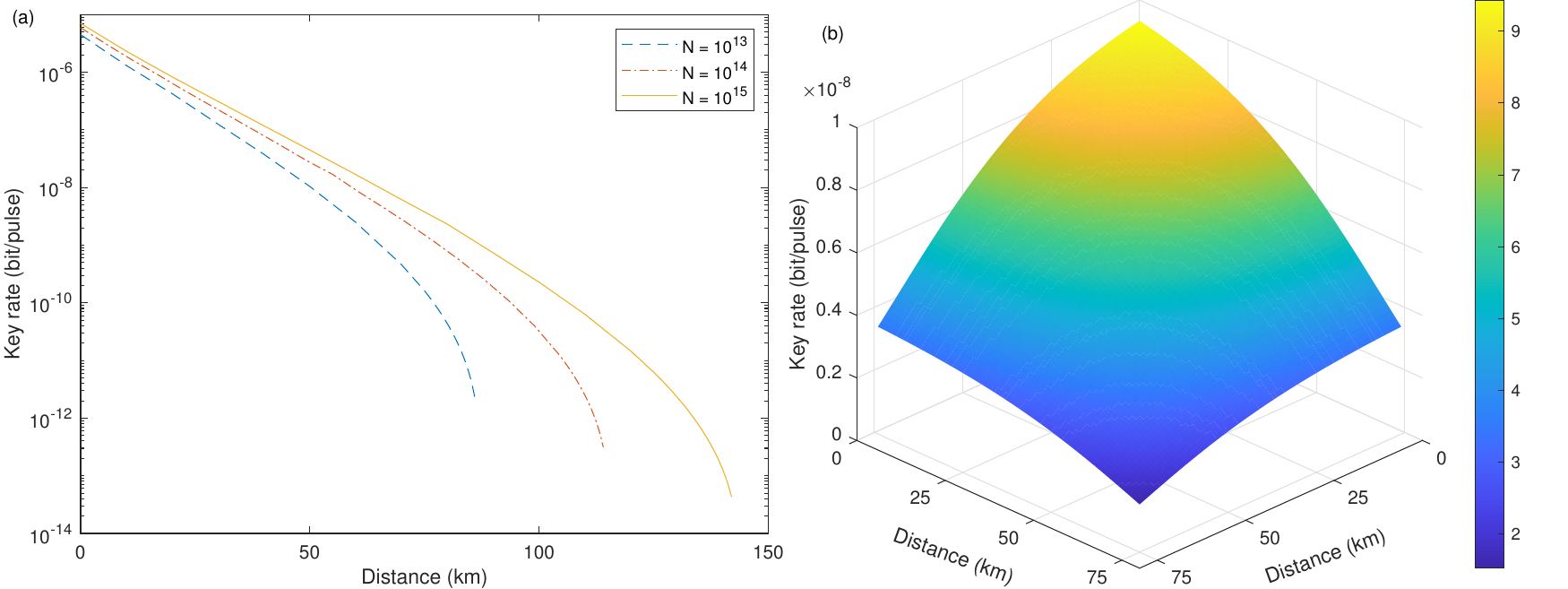}
		\caption{Simulation of key rates. The error correction efficiency $f=1.06$. The failure probability of bound for finite-size analysis $\epsilon=10^{-10}$. The number of phase slices $D=16$. The attenuation coefficient of optical fiber $\alpha = 0.175~\mathrm{dB/km}$. The detector dark count rate $p_d=2.4\times10^{-8}/\text{pulse}$. The total detection efficiency $\eta_d=0.6$. The intensity and sending probability of the coherent states prepared by each party are optimized to achieve the highest key rate. (a) Simulated key rates in symmetric fiber channels. The number of rounds $N$ corresponding to the dashed line, solid line, and dash-dotted line are $10^{13}$, $10^{14}$, and $10^{15}$, respectively. (b) Simulated key rates in asymmetric fiber channels. The distance from Alice to the measurement site is fixed at $\SI{75}{\km} $, while the distances from Bob and Charlie to the measurement site are represented by the X-axis and Y-axis, respectively. Here $N=4\times10^{13}$.}
	\label{fig:keyratesimu}
\end{figure*}

Finally, the key rates can be obtained by substituting Eq.~\eqref{eq:simulation_ex} and Eq.~\eqref{eq:simulation_Ez} into Eq.~\eqref{eq:keyrate}. The simulation results for symmetric and asymmetric channels are given in Fig.~\ref{fig:keyratesimu}. In Fig.~\ref{fig:keyratesimu} (a), the key rates in symmetric fiber channels are simulated under the number of rounds $N$ of $10^{13}$, $10^{14}$, and $10^{15}$, respectively. In Fig.~\ref{fig:keyratesimu} (b), the key rates in asymmetric fiber channels are simulated where the distance from Alice to the measurement site is fixed at $\SI{75}{\km} $, while the distances from Bob and Charlie to the measurement site are vary from 0 to $\SI{75}{\km} $, respectively.

\section{Experimental parameters and results} \label{sec:errd}
In this section, we provide more detailed experimental parameters and results for symmetric and asymmetric channels compared to the main text, as shown in Table~\ref{table:expprsupp}.

\begin{table*}[bpth!]
	\centering
	\caption{ Experimental parameters and results for symmetric and asymmetric channels. $N_{\mu}$ ($N_{\nu}$, $N_{\omega}$) is the number of rounds in which Alice, Bob, and Charlie simultaneously send signal (decoy) states with intensities $\mu_a$ ($\nu_a$, $\omega_a$), $\mu_b$ ($\nu_b$, $\omega_b$) and $\mu_c$ ($\nu_c$, $\omega_c$).  The probabilities of each party corresponding to the sending signal state and two decoy states are represented as $p_\mu$, $p_\nu$, and $p_\omega$, respectively. The transmittance from Alice (Bob, Charlie) to the measuremnt site is denoted as $\eta_A$ ($\eta_B$, $\eta_C$). Note that $\eta_A$ contains the loss of the BS used for the splitting of Alice's coherent pulse at the measurement site. $M_{\mu}$ ($M_{\nu}$, $M_{\omega}$) is the effective clicks for simultaneously sending signal (decoy) states with intensities $\mu_a$ ($\nu_a$, $\omega_a$), $\mu_b$ ($\nu_b$, $\omega_b$) and $\mu_c$ ($\nu_c$, $\omega_c$). $K$ is the raw key length. $R_1$ is the key rate considering the finite-size effects.  $R_2$ is the asymptotic key rate.} 
	\label{table:expprsupp}
     \scalebox{0.95}{
	\begin{tabular}{c|cccc|ccccc}
	\hline
	\hline
$L$ & $\{25,25,25\}$ & $\{50,50,50\}$ & $\{75,75,75\}$ & $\{100,100,100\}$ & $\{75,50,50\}$ & $\{75,25,25\}$ & $\{75,50,75\}$ & $\{75,25,75\}$ & $\{75,25,50\}$ \\
	\hline 
	$\mu_a$ & $7.39\times10^{-2}$ & $7.73\times10^{-2}$ & $7.21\times10^{-2}$ & $6.14\times10^{-2}$ & $1.05\times10^{-1}$ & $1.27\times10^{-1}$ & $8.45\times10^{-2}$ & $9.13\times10^{-2}$ & $1.17\times10^{-1}$\\
	$\mu_b$ & $3.75\times10^{-2}$ & $3.54\times10^{-2}$ & $3.57\times10^{-2}$ & $3.15\times10^{-2}$ & $1.85\times10^{-2}$ & $8.99\times10^{-3}$ & $1.50\times10^{-2}$ & $6.60\times10^{-3}$ & $8.18\times10^{-3}$\\
     $\mu_c$ & $3.58\times10^{-2}$ & $3.55\times10^{-2}$ & $3.42\times10^{-2}$ & $2.92\times10^{-2}$ & $1.90\times10^{-2}$ & $8.67\times10^{-3}$ & $4.14\times10^{-2}$ & $4.52\times10^{-2}$ & $2.02\times10^{-2}$\\
     $\nu_a$ & $3.88\times10^{-2}$ & $4.19\times10^{-2}$ & $3.95\times10^{-2}$ & $3.23\times10^{-2}$ & $5.74\times10^{-2}$ & $7.00\times10^{-2}$ & $4.63\times10^{-2}$ & $4.97\times10^{-2}$ & $6.40\times10^{-2}$\\
     $\nu_b$ & $1.95\times10^{-2}$ & $1.92\times10^{-2}$ & $1.96\times10^{-2}$ & $1.67\times10^{-2}$ & $1.01\times10^{-2}$ & $4.95\times10^{-3}$ & $8.31\times10^{-3}$ & $3.62\times10^{-3}$ & $4.47\times10^{-3}$\\
     $\nu_c$ & $1.87\times10^{-2}$ & $1.93\times10^{-2}$ & $1.87\times10^{-2}$ & $1.55\times10^{-2}$ & $1.04\times10^{-2}$ & $4.78\times10^{-3}$ & $2.30\times10^{-2}$ & $2.47\times10^{-2}$ & $1.10\times10^{-2}$\\
  $\omega_a$ & $1.88\times10^{-3}$ & $2.83\times10^{-3}$ & $2.88\times10^{-3}$ & $2.73\times10^{-3}$ & $3.96\times10^{-3}$ & $4.60\times10^{-3}$ & $3.39\times10^{-3}$ & $3.47\times10^{-3}$ & $4.31\times10^{-3}$\\
  $\omega_b$ & $9.79\times10^{-4}$ & $1.29\times10^{-3}$ & $1.46\times10^{-3}$ & $1.39\times10^{-3}$ & $7.18\times10^{-4}$ & $3.18\times10^{-4}$ & $5.98\times10^{-4}$ & $2.54\times10^{-4}$ & $3.03\times10^{-4}$\\
  $\omega_c$ & $9.53\times10^{-4}$ & $1.25\times10^{-3}$ & $1.42\times10^{-3}$ & $1.29\times10^{-3}$ & $7.16\times10^{-4}$ & $3.14\times10^{-4}$ & $1.61\times10^{-3}$ & $1.74\times10^{-3}$ & $7.48\times10^{-4}$\\
$N_{\mu}$	& $1.10\times10^{13}$ & $1.30\times10^{13}$ & $9.80\times10^{12}$ & $7.52\times10^{12}$ & $1.25\times10^{13}$ & $1.37\times10^{13}$ & $1.12\times10^{13}$ & $1.12\times10^{13}$ & $1.26\times10^{13}$\\
$N_{\nu}$	& $2.46\times10^{10}$ & $1.12\times10^{11}$ & $3.70\times10^{11}$ & $3.12\times10^{12}$ & $2.20\times10^{11}$ & $1.70\times10^{11}$ & $2.90\times10^{11}$ & $2.69\times10^{11}$ & $1.89\times10^{11}$\\
$N_{\omega}$ & $1.98\times10^{10}$ & $1.19\times10^{11}$ & $4.13\times10^{11}$ & $4.18\times10^{12}$ & $2.42\times10^{11}$ & $2.03 \times10^{11}$ & $3.33\times10^{11}$ & $3.16\times10^{11}$ & $1.95\times10^{11}$\\
	$p_\mu$ 	& 0.8003 & 0.7079 & 0.5950 & 0.3916 & 0.6572 & 0.6850 & 0.6258 & 0.6370 & 0.6706\\
	$p_\nu$ 	& 0.1030 & 0.1442 & 0.1977 & 0.2897 & 0.1685 & 0.1550 & 0.1833 & 0.1780 & 0.1621\\
	$p_\omega$	& 0.0967 & 0.1479 &0.2073 & 0.3187 &0.1743 &0.1600 & 0.1909 & 0.1850 & 0.1673\\
	\hline	
	$\eta_{A}$ 	& 0.1687 	& 0.0635 	& 0.0241 	& 0.0090 	& 0.0241 	& 0.0241 	& 0.0241 	& 0.0241 	& 0.0241\\
	$\eta_{B}$ 	& 0.3381 	& 0.1409 	& 0.0475	& 0.0171 	& 0.1409 	& 0.3381	& 0.1409 	&0.3381 	& 0.3381\\
	$\eta_{C}$ 	& 0.3420 	& 0.1380 	& 0.0499 	& 0.0194 	& 0.1380 	& 0.3420 	& 0.0499 	&0.0499 	& 0.1380\\

     \hline
$M_{\mu}$			&	2040135966 &	437817496	&	40466521	&	3557353	& 	115848639	& 	173509703	& 	66660580	&	77230873	&	135241099\\
$M_{\nu}$		& 	1272088 	& 	1102148	&	458942	&	410061	&	609615	&	659095 	&	526434	&	554995	&	604702\\
$M_{\omega}$	&	2482		&	5401		&	2790		&	3915		&	3306		&	3365		&	3045		&	3258		&	2872\\
     $K$				&	31944815 	&	6832717	&	631442	&	55251		&	1810617	&	2710929	&	1041251	&	1209357 	&	2116983\\
$R_1$ & $3.15\times10^{-7}$ & $1.79\times10^{-8}$ & $1.32\times10^{-9}$ & $4.20\times10^{-11}$ & $4.73\times10^{-9}$ & $6.83\times10^{-9}$ & $2.90\times10^{-9}$ & $3.00\times10^{-9}$ & $6.07\times10^{-9}$\\
$R_2$ & $8.37\times 10^{-7}$ & $8.28\times 10^{-8}$ & $9.26\times 10^{-9}$ & $9.89\times 10^{-10}$ & $3.01\times 10^{-8}$ & $3.82\times 10^{-8}$ & $2.13\times 10^{-8}$ & $2.15\times 10^{-8}$ & $3.57\times 10^{-8}$\\
	\hline    
	\hline    	
	\end{tabular}}
\end{table*}
\bibliographystyle{apsrev4-2}

%

\end{document}